\documentstyle[amscd,amssymb,verbatim,12pt]{amsart}
\def\dspace{\baselineskip=0.3 in}
\begin{document}
\dspace
\title[Curvature Inspired Cosmological Scenario]{CURVATURE INSPIRED COSMOLOGICAL SCENARIO }

\author{\bf S.K.Srivastava}
{ }
\maketitle
\centerline{ Department of Mathematics,}
 \centerline{ North Eastrn Hill University,}
 \centerline{  Shillong-793022, India}
\centerline{ srivastava@@.nehu.ac.in; sushilsrivastava52@@gmail.com}

\vspace{1cm}

\centerline{\bf Abstract}

Using modified gravity with non-linear terms of curvature , $R^2$ and
$R^{(2 + r)}$ (with $r$ being a positive real number and $R$ being the scalar
curvature), cosmological scenario, beginning at the Planck scale, 
is obtained. Here a unified picture of cosmology is obtained from
$f(R)-gravity$. In this scenario, universe begins with power-law inflation
followed by deceleration and acceleration in the late universe as well as
possible collapse of the unverse in future. It is different from
$f(R)-${\em dark energy models} with non-linear curvature terms  assumed as
dark energy.  Here,  dark energy terms
 are induced by linear as well as non-linear terms of curvature in Friedmann equation being
 derived from modified gravity .  It is also interesting to see that, in this
 model, dark 
radiation and dark matter terms emerge spotaneously from the gravitational
sector. 
It is found that 
dark energy, obtained here, behaves as quintessence in the early universe and
phantom in the late universe.
 Moreover, analogous to
brane-tension in brane-gravity inspired Friedmann equation, a tension term  $\lambda$
 arises here being called as cosmic tension,
 It is found that, in the late universe, Friedmann equation (obtained here)  contains a term $- \rho^2/2\lambda$ ($\rho$ being the phantom energy
density) analogous to a similar term in Friedmann equation with loop quantum
effects, if $\lambda > 0$ and brane-gravity correction when  $\lambda < 0$.

\vspace{1cm}

\centerline {\underline{\bf 1. Introduction}}
 \smallskip

Cosmology was revolutionized by observations made during last few years
\cite{sp, ag}. These observations show conclusive evidence for acceleration in
the late universe, which is still a challenge for cosmologists. Theoretically,
it is found that dark energy (DE) violating 
{\em strong energy condition}(SEC) or {\em weak energy condition}(WEC) is
responsible for it. So, in the recent past, many DE models were proposed to
explain the {\em late cosmic acceleration}. A comprehensive review of these
models is available in \cite{ejc}. Later on, it was realized that even
non-linear terms of curvature $R^{-n}$ with $n >0$ also could be used as DE
\cite{cap}. Although this model explained late cosmic acceleration, it
exhibited instability and failed to satisfy solar system constraints. It was
improved further by Nojiri and Odintsov taking different forms of $f(R)$ for
DE. These improved models satisfied solar system constraints exhibiting late
cosmic acceleration for small curvature and early inflation for large
curvature. Thus, in $f(R)-$ {\em dark energy models},  non-linear curvature
terms are considered as an alternative for DE \cite[for detailed
review]{snj}.  Recently, in \cite{lds}, it is shown that $f(R)-$ {\em dark
  energy models} with dominating powers of $R$
 for large or small $R$ can not yield viable cosmolgy as results contradict the
 standard model and do not satisfy Wilkinson Microwave Anisotropy
 Probe (WMAP) results, though these
 models pass solar system constraints and explain late acceleration. Moreover,
 it is also shown that
 the most popular model with $f(R) = R + \alpha R^m + \beta R^{-n} (m >0, n
 >0)$ considered in \cite{snj}
 is unable to  produce matter in the late universe prior to the beginning of
 late acceleration  \cite{lds}.

In what follows, it is aimed to get a viable cosmolgy consistent with WMAP
from $f(R)- gravity,$ where $f(R) = R/16\pi G + $ powers of $R$,
{\em not} from  $f(R)-$ {\em dark energy models} \cite{sks, sks06, sks07}
discussed above . There is a  crucial
difference between the two. In the latter case, which is
criticised in \cite{lds}, non-linear terms of curvature are treated as dark
energy terms. On the contrary, in the former case, neither linear nor
non-linear term is considered as dark energy. In the present model, it is
important to see that DE terms
are induced by linear
(Einstein-Hilbert term) as well as  non-linear terms $R^2$ and $ R^{(2 + r)}$ in the action. In $f(R)-$ {\em dark energy models},
dark energy terms depend on $f(R)$ terms and its derivative $F = df/dR$. In
the former case, induced DE terms depend on the scale factor $a(t)$ of the
homogeneous and flat of Friedmann - Robertson - Walker (FRW) model of the
universe.  

In this paper, the $f(R)-gravity$ based modified Friedmann equations are
derived in the early and late
universe taking small and large $a(t)$ respectively. Contrary
to  $f(R)-$ {\em dark energy models} dominance of non-linear terms of
curvature is not taken here for small and large $R$ as, in the present model,
DE terms emerge as imprints of both linear term as well as powers of $R$. In
$f(R)-$ {\em dark energy models}, action contains lagrangians for
matter and radiation \cite{snj, lds}. It is interesting to see that, in the
present $f(R)- gravity$ model of cosmology, dark radiation and dark matter
terms emerge spontaneously \cite{sks06, sks07}.

 Friedmann equation (FE), obtained here, gives cosmic
 dynamics, where some 
 terms emerge  having forms different from known forms of energy
 density (radiation or matter) and violate SEC in the case of early universe
 and WEC in the case of late universe. So, these terms are recognized as
 curvature induced DE.
  Thus, this approach of getting DE from
 curvature is {\em different} from the approach of \cite{snj, lds}. It is also
 interesting to see that, in the case of late universe, FE contains square of
 DE density with negative sign analogous to a similar term in Friedmann
 equation with loop quantum effects \cite{ms}.

 In what follows, it is
interesting to see that radiation and matter terms in FE (obtained here)
emerge spontaneosly 
if $r = 3$ and $n = 1/4$. In \cite{snj}, radiation and and matter terms do not
emerge from the gravitational sector. It is important to mention that, here,
theory itself ignores the cases of $r =1,2$ i.e. theory suggests that
non-linear of $R$ should be $R^2$ and 
$R^5$. In \cite{snj}, a certain form of scale factor $a(t)$ is assumed and, later on,
conditions are obtained for assumed $a(t)$  consistent with
experiments and satisfying stability criterion. Here, $a(t)$ is {\em not}
assumed, but it is derived solving Friedmann equations in the different stages
of the universe. This is another important {\em difference} in approach of
this paper
compared to \cite{snj}.  All these results are obtained from
modified gravity without using any exotic matter or field. This approach is
adapted in \cite{sks,sks06, sks07} also.

In \cite{sks07}, a unified picture of the universe, from early inflationary
stage to late acceleration and deceleration driven by dark radiation and dark
matter between these two stages,  is obtained taking linear and non-linear
terms of curvature as well as a different scalar. But, in this paper, the same
result is obtained from curvature terms only .

Investigations, given below, show that, originating at the Planck scale,  the
universe inflates for a very short period followed by deceleration
  driven by curvature-induced dark radiation and
subsequently 
by dark matter. In the very late universe, around $12.86 {\rm Gyr}$, a transition
from {\em deceleration} to {\em acceleration} takes place. Further, it is
found that late acceleration will continue upto $58.48 {\rm Gyr}$. This is an
epoch for transition from {\em acceleration} to {\em deceleration}. At this
epoch, acceleration will stop and deceleration, driven by matter, will resume
\cite{sks07, vs}. The decelerated expansion will continue upto the time $\sim
1.05 \times 10^{156} {\rm Gyr}$. By this time, the universe will have maximum
expansion. So, it is natural to think that the universe will retrace back and
contract. Results show that, due to contraction, universe will  collapse by the
time $\sim 1.3 \times 10^{156} {\rm Gyr}$. 

 In \cite{bo}, quintessence DE
in the early universe and phantom DE in the late universe have been considered
taking non-gravitational scalar field (curvature independent scalar field) as
DE source. The present paper is different from \cite{bo} in the sense that here
we have gravitational origin of quintessence and phantom DE as it is obtained
in references \cite{sks,sks06, sks07}.  

Natural units $(k_B = {\hbar} = c = 1)$ (where $k_B, {\hbar}, c$ have their
usual meaning. GeV is used as a fundamental unit and we have $1 {\rm GeV}^{-1}
= 6.58 \times 10^{-25} sec.$ and $1 {\rm GeV} = 1.16 \times 10^{13} K.$

\bigskip

\centerline {\underline{\bf 2. Action for $f(R)- gravity$ and Friedmann
    equations }}

 \smallskip

 Here action is taken as
$$ S = \int {d^4x} \sqrt{-g} \Big[\frac{R}{16 \pi G} + \alpha R^2 + \beta
R^{(2 + r)}  \Big],  \eqno(2.1)$$ 
where $G = M_P^{-2} (M_P = 10^{19} {\rm GeV}$ is the Planck mass), $\alpha$ is
a dimensionless coupling constant, $\beta$ is a constant having dimension
(mass)$^{(-2 r)}$ (as $R$ has mass dimension 2) with $r$ being a positive real
number.

Using the condition $\delta S/\delta g^{\mu\nu} = 0$, action (2.1) yields field equations
$$ \frac{1}{16 \pi G} (R_{\mu\nu} - \frac{1}{2} g_{\mu\nu} R) + \alpha (2
\triangledown_{\mu} \triangledown_{\nu} R - 2 g_{\mu\nu} {\Box} R -
\frac{1}{2} g_{\mu\nu} R^2 + 2 R R_{\mu\nu} ) $$
$$ + \beta (2 + r) ( \triangledown_{\mu} \triangledown_{\nu} R^{(1 + r)} -
g_{\mu\nu} {\Box} 
R^{(1 + r)}) + \frac{1}{2}\beta g_{\mu\nu} R^{(2 + r)} $$
$$- \beta (2 + r) R^{(1 +   r)} R_{\mu\nu}   = 0, \eqno(2.2)$$  
where $\triangledown_{\mu}$ stands for the covariant derivative.

Taking trace of (2.2), it is obtained that
$$ - \frac{R}{16 \pi G} - 6 \alpha{\Box} R - 3 \beta (2 + r) {\Box} R^{(1 +
 r)}  + \beta r R^{(2 + r)}  = 0 \eqno(2.3)$$
with
$$ {\Box} = \frac{1}{\sqrt{-g}} \frac{\partial}{\partial x^{\mu}}
\Big(\sqrt{-g} g^{\mu\nu} \frac{\partial}{\partial x^{\nu}} \Big). \eqno(2.4)$$

In (2.3)
$${\Box} R^{(1 + r)} = (1 + r) [R^{r}{\Box} R + r
R^{(r - 1)} \triangledown^{\mu}R \triangledown_{\mu}R ]. \eqno(2.5)$$ 

From (2.3) and (2.5)
$$ - \frac{R}{16 \pi G} - [6 \alpha + 3 \beta (1 + r)(2 + r)  R^r ] {\Box} R -
3 \beta r (1 + r)(2 + r)R^{(r - 1)} \triangledown^{\mu}R \triangledown_{\mu}R$$
$$+ \beta r R^{(2 + r)}  = 0 \eqno(2.6)$$ 

In (2.6), $[6 \alpha + 3 \beta (1 + r)(2 + r)  R^r ]$
emerges as a coefficient of ${\Box} R$ due to 
presence of terms $\alpha R^2$ and $\beta R^{(2 + r)}$ in the action
(2.1). If $\alpha = 
0$, effect of $R^2$ vanishes and  effect of $R^{(2 + r)}$ is switched
off for $\beta = 0$. So, like \cite{sks07} an {\em effective} scalar curvature
${\tilde R}$ is defined as 
$$ \gamma {\tilde R}^r = [6 \alpha + 3 \beta (1 + r)(2 + r)  R^r ] , \eqno(2.7)$$
where  $\gamma$ is a constant having dimension (mass)$^{-2r}$ being
used for dimensional correction.

Using (2.7) in (2.6), we have
$$\frac{1}{16 \pi G} Y^{1/r} - (\gamma/r) {\tilde R}^r Y^{(1/r - 2)} [Y {\Box}
Y + (1/r - 1) 
\triangledown^{\mu} Y \triangledown_{\mu} Y ]$$ 
$$ - 3 (\beta/r) (1 + r)(2 + r) Y^{(1/r - 1)} \triangledown^{\mu} Y \triangledown_{\mu} Y  + \beta r Y^{(2 + r)/r} = 0, \eqno(2.8)$$
where
$$ Y = R^{r} = \frac{ \gamma {\tilde R}^{r} - 6 \alpha}{3 \beta (1 + r)(2 + r)}. \eqno(2.9a)$$
(2.8) is simplified as 
 $$\frac{1}{16 \pi G} Y - (\gamma/r) {\tilde R}^{r}[{\Box} Y + (1/r - 1)Y^{-1} \triangledown^{\mu} Y \triangledown_{\mu} Y ]$$
$$ - 3 (\beta/r) (1 + r)(2 + r)  \triangledown^{\mu} Y \triangledown_{\mu} Y + \beta r Y^{(1/r + 2)} = 0. \eqno(2.9b) $$

Using (2.9a)in (2.9b), it is obtained that
$$-\frac{r}{16 \pi G \gamma} \Big[\frac{6 \alpha}{\gamma {\tilde R}^r} - 1 \Big] + {\Box}{\tilde R}^r - (1/r - 1) \frac{\gamma}{[6 \alpha - \gamma {\tilde R}^{r}]} \triangledown^{\mu}{\tilde R}^{r} \triangledown_{\mu}{\tilde R}^{r} + {\tilde R}^{- r} \triangledown^{\mu}{\tilde R}^{r}\triangledown_{\mu}{\tilde R}^{r}$$
$$ + [3 \beta^2 r (1 + r)(2 + r)/ \gamma^2] {\tilde R}^{r} \Big[\frac{\gamma {\tilde R}^{r}- 6 \alpha}{ 3 \beta (1 + r)(2 + r)} \Big]^{(1/r + 2)} = 0 . \eqno(2.10)$$

(2.10) is re-written as
$$-\frac{1}{16 \pi G} \frac{1}{\gamma {\tilde R}^{r - 1}}\Big[\frac{6
  \alpha}{\gamma {\tilde R}^{r}} - 1 \Big] + {\Box}{\tilde R} +
  (r - 1) {\tilde R}^{-1} \triangledown^{\mu}{\tilde R}
  \triangledown_{\mu}{\tilde R}$$
$$ - (1 - r) \frac{\gamma {\tilde R}^{r - 1}}{6 \alpha - \gamma {\tilde R}^{r}}\triangledown^{\mu}{\tilde R}\triangledown_{\mu}{\tilde R} +
  r {\tilde R}^{-1} \triangledown^{\mu}{\tilde R} 
  \triangledown_{\mu}{\tilde R}$$
$$+ r)(2 + r)/ \gamma^2]
  {\tilde R}^{2 r - 1} \Big[\frac{\gamma {\tilde R}^{r} - 6\alpha}{3 \beta (1 + r)(2 + r)} \Big]^{(1/r + 2)} = 0 . \eqno(2.11)$$ 

Experimental evidences \cite{ad} support spatially homogeneous
flat model of the universe 
$$dS^2 = dt^2 - a^2(t) [dx^2 + dy^2 + dz^2] \eqno(2.12)$$
with $a(t)$ being the scale factor.

For $a(t)$, being the power-law function of cosmic time, ${\tilde R} \sim a^{-n}$. For
example,  ${\tilde R} \sim a^{-3}$ for
matter-dominated model. So, there is no harm in taking
$$  {\tilde R} = \frac{A}{ a^n} , \eqno(2.13)$$
where $n > 0$ is a real number and $A$ is a constant with mass dimension 2. 

Connecting (2.11) and (2.13), it is obtained that
$$ \frac{\ddot a}{a} + \Big[2 - n - n (r - 1) + \frac{n (1 - r) \gamma A^r
  a^{- nr}}{6 \alpha - \gamma A^r a^{- nr}} - n r \Big] \Big(\frac{\dot a}{a} \Big)^2 = \frac{a^{n r}}{16 \pi G \gamma A^r} \Big[\frac{6 \alpha a^{nr}}{\gamma A^r} - 1 \Big]$$
$$ - \frac{ \beta^{-1/3} }{n (\gamma A^r a^{-nr})^2[3r(1 + r)(2 + r)]^{1 + 1/r} }[6 \alpha - \gamma A^r a^{-nr}]^{2 + 1/r} , \eqno(2.14)$$
taking $(- \beta)^{-1/3} = - \beta^{-1/3}$ and ignoring complex roots as these roots lead  to unphysical situations. Now, we have follwing two cases.

\smallskip

\noindent {\bf Case 1 : The Early Universe}

In this case, $a(t)$ is very small, so (2.14) is approximated as
$$ \frac{\ddot a}{a} + \Big[2 - n - n r \Big] \Big(\frac{\dot a}{a} \Big)^2 \simeq - \frac{\beta^{-1/3} }{n (\gamma A^r a^{-nr})^2[3r(1 + r)(2 + r)]^{1 + 1/r} }$$
$$\times[6 \alpha - \gamma A^r a^{-nr}]^{2 + 1/r}  \eqno(2.15)$$
as
$$\frac{ \gamma A^r
  a^{- nr}}{6 \alpha - \gamma A^r a^{- nr}} \approx - 1.\eqno(2.16)$$

Integration of (2.16) leads to
$$ \Big(\frac{\dot a}{a} \Big)^2 = \frac{B}{a^{(2 + 2M)}} - \frac{2 \beta^{-1/r}}{n (\gamma A^r)^2 [3r(1 + r)(2 + r)]^{1 + 1/r}a^{(2 + 2M)}}$$
$$\times \int a^{(1 + 2M + 2nr)}[6 \alpha - \gamma A^r a^{-nr}]^{2 + 1/r}  \eqno(2.17)$$
with
$$ M = 2 - n - n r. \eqno(2.18)$$

\smallskip

\noindent {\bf Case 2 : The Late Universe}

In this case, $a(t)$ is large, so (2.14) is approximated as
$$\frac{\ddot a}{a} + \Big[2 - n - n (r - 1)- nr \Big] \Big(\frac{\dot a}{a} \Big)^2 \simeq  \frac{a^{n r}}{16 \pi G \gamma A^r} \Big[\frac{6 \alpha a^{nr}}{\gamma A^r} - 1 \Big] $$
$$ - \frac{ \beta^{-1/3} }{n (\gamma A^r a^{-nr})^2[3r(1 + r)(2 + r)]^{1 + 1/r} }(6\alpha)^{2 + 1/r} [a^{2 nr} - (2 + 1/r)\gamma A^r a^{nr}] \eqno(2.19a)$$
as 
$$\frac{ \gamma A^r
  a^{- nr}}{6 \alpha - \gamma A^r a^{- nr}} \approx 0$$
for large scale factor $a$. So, (2.19a) is re-written as
$$\frac{\ddot a}{a} + \Big[2 - 2nr \Big] \Big(\frac{\dot a}{a} \Big)^2 = D a^{nr} - E a^{2nr} , \eqno(2.19b)$$
where
$$D = \Big(\frac{6\alpha}{\gamma A^r}\Big) \Big[ \frac{1}{16 \pi G n} - (2 +
1/r)\frac{[3 r (1 + r)(2 + r)]^{-1-1/r}}{n}\Big(\frac{6\alpha}{\gamma
  A^r}\Big) \Big] \eqno(2.20a)$$  
and
$$E = \Big(\frac{6\alpha}{\gamma A^r}\Big)^2 \Big[ \frac{1}{16 \pi G n} -
\frac{[3 r (1 + r)(2 + r)]^{-1-1/r}}{n}\Big(\frac{6\alpha}{\gamma A^r}\Big)
\Big]. \eqno(2.20b)$$  

(2.19b) is integrated to
$$ \Big(\frac{\dot a}{a} \Big)^2 = \frac{B}{a^{(2 + 2N)}} + \frac{2D}{(2 + 2N +
    nr)} a^{nr} \Big[1 - \frac{E(2 + 2N + nr)}{D(2 + 2N + 2nr)}a^{nr} \Big]
  \eqno(2.21)$$  
with
$$ N = 2 - 2nr .\eqno(2.22)$$ 

Further, it is found that if $M = 1$, the first term on r.h.s.(right hand
side) of (2.17)gives radiation. Moreover, if $N = 1/2$  the first term of r.h.s. of
(2.21) gives matter. So, using $M = 1$ in (2.18) and $N = 1/2$ in (2.22), it is
obtained that 
$$ nr = \frac{3}{4}, \eqno(2.23)$$
$$ n = \frac{1}{4}\eqno(2.24)$$
and
$$ r = 3. \eqno(2.25)$$

\bigskip
\bigskip

\centerline {\underline{\bf 3. Power-law inflation followed by deceleration }} 

\centerline {\underline{\bf in the early universe }}

 \smallskip

 The approximated Friedmann equation (2.17), in the case of the early universe, looks like
$$ \Big(\frac{\dot a}{a} \Big)^2 = \frac{B}{a^4} - \frac{8 \beta^{-1/3}}{ (\gamma A^3)^2 [180]^{4/3}a^4} \int a^{9/2}[6 \alpha - \gamma A^3 a^{- 3/4}]^{7/3}  \eqno(3.1)$$
using definitions of $M$ and $N$ as well as (2.24) and (2.25). In (3.1),
$$\int a^{9/2}[6 \alpha - \gamma A^3 a^{- 3/4}]^{7/3} = \Big[\frac{2}{11} a^{11/2}\{6 \alpha - \gamma A^3 a^{- 3/4}\}^{7/3}\Big] - \frac{7}{22} \gamma A^3$$
$$\times \int a^{15/4}\{6 \alpha - \gamma A^3 a^{- 3/4}\}^{4/3} da .\eqno(3.2a)$$
It is noted that for
$$a < \Big(\gamma A^3/6 \alpha \Big)^{4/3} = a_c , \eqno(3.2b)$$
terms within bracket and the integral on the right hand side of (3.2a) are of
the order of $a^{15/4}$.

So,
$$\int a^{9/2}[6 \alpha - \gamma A^3 a^{- 3/4}]^{7/3} \approx [\frac{2}{11}
a^{11/2}\{6 \alpha - \gamma A^3 a^{- 3/4}\}^{7/3} .\eqno(3.2c)$$ 
Thus, using (3.2a,b,c),(3.1) is approximated as
$$ \Big(\frac{\dot a}{a} \Big)^2 \approx \frac{B}{a^4} - \frac{16
  \beta^{-1/3}}{11 (\gamma A^3)^{-1/3} [180]^{4/3}}  a^{3/2}\Big[a_c^{-3/4} - a^{- 3/4}\Big]^{7/3}  \eqno(3.3)$$ 

It is interesting to see that a radiation density  term $B/a^4$ emerges
spontaneously. This type of a term, being called dark radiation, emerges in brane-gravity inspired Friedmann
equation too. So, analogous to brane-gravity, here also $B/a^4$ is called dark radiation. Other terms on r.h.s. of (3.3) are caused by linear as well as non-linear
terms of 
curvature in the action (2.1). These terms also constitute energy density term
$$ \rho^{\rm qu}_{\rm de} =  \frac{3}{8 \pi G} \Big[\frac{16
  \beta^{-1/3}}{11 (\gamma A^3)^{-1/3} [180]^{4/3}}  a^{3/2}\Big]\Big[ a^{- 3/4} -
  \frac{6 \alpha}{\gamma A^3}\Big]^{7/3}  \eqno(3.4)$$
(taking real root of $(-1)^{-1/3}$ as above) satisfying the conservation
  equation 
$$ {\dot \rho}_{\rm de} + 3 \frac{\dot a}{a} ( \rho_{\rm
  de} + p_{\rm de} ) = 0. \eqno(3.5)$$

Connecting (3.4) and (3.5),  equation of state (EOP
is obtained as

$$p^{\rm qu}_{\rm de} =  - \frac{3}{2}\rho^{\rm qu}_{\rm de} + \frac{7}{12} f [a^{-3/4} -
a_c^{-3/4}]^{4/3}, \eqno(3.6a)$$
where
$$ f = \frac{3}{8\pi G}  \frac{16
  \beta^{-1/3}}{11 (\gamma A^3)^{-1/3} [180]^{4/3}}  \eqno(3.6b)$$
(3.6a) is the scale factor-dependent equation of state parameter,valid for $a_P
\le a(t) < a_c$. Such an equation of state parameter is obtained in
\cite{sks06} also. It yields
$$\rho^{\rm qu}_{\rm de} + p^{\rm qu}_{\rm de} > o \quad {\rm and}\quad
\rho^{\rm qu}_{\rm de} + 3 p^{\rm qu}_{\rm de} < o ,$$
for $a_P \le a(t) \le a_c$.
It shows that DE, having energy density (3.4) mimics {\em quintessence dark
  energy} \cite{sks, sks06,  sks07}.

Here investigations start at the Planck scale, where DE density is obtained
around $10^{75} {\rm GeV}^4$. So, (3.4) is obtained as
$$\rho^{\rm qu}_{\rm de} = F a^{3/2} [ a^{- 3/4} - a_c^{- 3/4}]^{7/3}
\eqno(3.7 a)$$
with
$$ F =  10^{75} {a_P^{-3/2}\Big[ a_P^{- 3/4} - a_c^{- 3/4} 
  \Big]^{-7/3}}.\eqno(3.7b)$$
Thus , (3.6b) and (3.7b) imply
$$ f = F .\eqno(3.7c)$$

Connecting (3.3) and (3.7a), it is obtained that
$$ \Big(\frac{\dot a}{a} \Big)^2 \simeq \frac{B}{a^4} + \frac{8 \pi \times
  10^{37}}{3} \Big(\frac{ a}{a_P}\Big)^{3/2} \Big[\frac{ a^{- 3/4} - a_c^{-
  3/4}  }{ a_P^{- 3/4} - a_c^{- 3/4}}
  \Big]^{7/3} \eqno(3.8)$$ 
using $ G = M_P^{-2} = 10^{-38} {\rm GeV}^{-2}.$

It means that the universe is driven by radiation for $a \ge a_c$. Moreover,
(3.7a) shows that $\rho^{\rm qu}_{\rm de}$ vanishes at $a = a_c$ and for $a_P <
a(t) < a_c,$ cosmic dynamics is given by
\begin{eqnarray*}
 \Big(\frac{\dot a}{a} \Big)^2 &\simeq & \frac{8 \pi \times
  10^{37}}{3} \Big(\frac{ a}{a_P}\Big)^{3/2} \Big[\frac{ a_c^{- 3/4} - a^{-
  3/4} }{ a_c^{- 3/4} - a_P^{- 3/4}}  \Big]^{7/3} \\ &\simeq & \frac{8 \pi \times
  10^{37}}{3} \Big(\frac{ a}{a_P}\Big)^{-1/4}. 
\end{eqnarray*}
\vspace{-1.2cm}
\begin{flushright}
(3.9) 
\end{flushright}
\smallskip

(3.9) integrates to
$$ a(t) = a_P \Big[1 + 10^{18}\sqrt{\frac{5 \pi}{12}}(t - t_P) \Big]^8 \eqno(3.10)$$
showing {\em acceleration} as ${\ddot a} > 0.$ 

If expansion (3.10) yields sufficient inflation in the early universe, 
$$ \frac{a_c}{a_P} = 10^{28}. \eqno(3.11)$$
The universe comes out of the inflationary phase at $t = t_c$ when $a(t)$
acquires the value $a_c$. So, from (3.10) and (3.11), it is obtained that
$$
 t_c \simeq  t_P +  10^{-18} \sqrt{\frac{12}{5 \pi}}\Big[\Big(\frac{a_c}{a_P} \Big)^{1/8}- 1
 \Big] \simeq   2.76 \times10^{4} t_P  \eqno(3.12)$$
using (3.11).

For $a \ge a_c,$ we have Friedmann equation (3.8) as
$$ \Big(\frac{\dot a}{a} \Big)^2 = \frac{B}{a^4}.\eqno(3.13)$$
This equation integrates to 
$$ a(t) = a_c [1 + \sqrt{B} (t - t_c)]^{1/2}. \eqno(3.14)$$
 (3.14) yields ${\ddot a} < 0$ showing {\em   deceleration} driven by dark
 radiation term.

\bigskip
\bigskip

\centerline {\underline{\bf 4. Deceleration followed by acceleration  in
 the late universe}}

\centerline {\underline{\bf as well as future collapse of the universe}}
 \smallskip

 In the late universe, the effective
Friedmann equation is given by (2.21). Using (2.23)-(2.25) in (2.21), we
obtain 
$$ \Big(\frac{\dot a}{a}\Big)^2 = \frac{C}{a^3} + \frac{8 D}{15} a^{3/4}
\Big[1 - \frac{5 E}{6 D} a^{3/4} \Big] , \eqno(4.1a)$$
where
$$ D = \frac{a_c^{-3/4}}{4 \pi G}\Big[1 - \frac{28 \pi G}{135}
\Big(\frac{1}{180 \beta} \Big)^{1/3} a_c^{-1/4} \Big] \eqno(4.1b)$$ 
and
$$ E = \frac{a_c^{-3/2}}{4 \pi G}\Big[1 - \frac{4 \pi G}{135}
\Big(\frac{1}{180 \beta} \Big)^{1/3} a_c^{-1/4} \Big] \eqno(4.1c)$$ 
being obtained from (2.20a) and (2.20b) using (2.23)-(2.25) and (3.2b).

The first term, on r.h.s. of (4.1a), emerges spontaneously and has the form of
matter density, so it is recognized as dark matter density like dark
radiation. Moreover, the second and third terms on r.h.s. of (4.1a) emerges
due to linear and non-linear terms of curvature. It is interesting to see that if
$$ \rho^{\rm ph}_{\rm de} = \frac{D}{5 \pi G} a^{3/4} \eqno(4.2)$$
and
$$\lambda = \frac{3 D^2}{25 \pi GE},  \eqno(4.3)$$
(4.1a) looks like
$$ \Big(\frac{\dot a}{a}\Big)^2 = \frac{8 \pi G}{3} \Big[\frac{ 3 C}{8 \pi G a^3} + \rho^{\rm ph}_{\rm de}
\Big\{1 - \frac{\rho^{\rm ph}_{\rm de}}{2 \lambda} \Big\}\Big]  \eqno(4.4)$$

Conservation equation (3.5) for $\rho^{\rm ph}_{\rm de}$ yields
$$ {\rm w}^{\rm ph}_{\rm de} = - \frac{5}{4}. \eqno(4.5)$$
(4.5) shows that the curvature-induced energy density $\rho^{\rm ph}_{\rm de}$
mimcs phantom dark energy as $ {\rm w}^{\rm ph}_{\rm de} < - 1.$ Thus, in the
late universe, a
phantom model is obtained from curvature without using any source of exotic
matter. Apart from this, (4.4) contains a term
$-{(\rho^{\rm ph}_{\rm de})^2}/{2 \lambda}$ analogous to brane-gravity
correction to the Friedmann equation (FE) for negative brane-tension \cite{rm} and
modifications in FE due to loop-quantum effects \cite{ms}. Here $\lambda$ is
called {\em cosmic tension} \cite{sks, sks06, sks07}.

According to WMAP results \cite{abl}, present density of pressureless dark
matter is obtained to be $\rho^{(m)}_0 = 0.23\rho_0^{\rm cr}$ and present dark
energy density $\rho^{\rm ph}_{{\rm de}0} = 0.73\rho_0^{\rm cr}$      with
$$\rho_0^{\rm cr} = \frac{3 H_0^2}{8 \pi G},$$
where current Hubble's rate of expansion $H_0 = 100h km/Mpc second = 2.32
\times 10^{-42} h {\rm GeV}$ and $h = 0.68$. Thus,
$$ \rho_0^{\rm cr} = 2.9 \times 10^{-47} {\rm GeV}^4. \eqno(4.6)$$

Using these values, it is obtained that
$$ \rho^{(m)} = \frac{ 3 C}{8 \pi G a^3} = \frac{6.67 \times 10^{-48}}{a^3} . \eqno(4.7)$$
and
$$ \rho^{\rm ph}_{\rm de} = \frac{D}{5 \pi G} a^{3/4} = {2.117 \times 10^{-47}}{a^{3/4}}  \eqno(4.8)$$
from (4.2).

Connecting (4.4), (4.7) and (4.8), it is obtained that
$$ \Big(\frac{\dot a}{a}\Big)^2 = \frac{8 \pi G}{3} \Big[\frac{6.67 \times
  10^{-48}}{a^3} + 2.12 \times 10^{-47} a^{3/4}
\Big\{1 - \frac{2.117 \times 10^{-47} a^{3/4}}{2 \lambda} \Big\}\Big]  \eqno(4.9)$$

(4.9) shows that
$$\frac{6.67 \times
  10^{-48}}{a^3} > 2.117 \times 10^{-47} a^{3/4}$$
for $ a < 0.735$ and 
$$\frac{6.67 \times
  10^{-48}}{a^3} < 2.117 \times 10^{-47} a^{3/4}$$
for $ a > 0.735.$

It means that a {\em transition} from matter-dominance to DE-dominance takes place at
$$ a_* = 0.735\eqno(4.10)$$
giving {\em red-shift}
$$ z_* = \frac{1}{a_*} - 1 = 0.3607 \eqno(4.11)$$
which is very closed to lower limit of $z_*$ given by 16 Type supernova observations \cite{ag}.
Thus, for $ a < 0.735$, (4.9) is approximated as

$$ \Big(\frac{\dot a}{a}\Big)^2 = \frac{8 \pi G}{3} \Big[\frac{6.67 \times
  10^{-48}}{a^3} \Big]= \frac{5.59 \times
  10^{-85}}{a^3} , \eqno(4.12)$$
which integrates to
$$a(t) = a_d [1 + 7.48\times 10^{-43} a_d^{-3/2}(t - t_d)]^{2/3} . \eqno(4.13)$$
It shows decelerated expansion as ${\ddot a} < 0.$

When $ a \ge 0.735$, (4.9) is approximated as
$$ \Big(\frac{\dot a}{a}\Big)^2 = 1.77 \times 10^{-84} a^{3/4}
\Big[1 - \frac{2.117 \times 10^{-47} a^{3/4}}{2 \lambda} \Big] . \eqno(4.14)$$

(4.14) integrates to
$$ a(t) = \Big[ \frac{2.117\times 10^{-47}}{2 \lambda} + \Big\{\sqrt{a_*^{-3/4} - \frac{2.117\times 10^{-47}}{2 \lambda}} - 5 \times 10^{-43} (t - t_*) \Big\}^2 \Big]^{-4/3}.\eqno(4.15)$$ 

This scale factor yields ${\ddot a} > 0$ showing {\em acceleration} in the late universe. In (4.15), $t_*$ is the time of transition from deceleration to acceleration in the late universe.

Using the present age of the universe $t_0 = 13.7 {\rm Gyr} = 6.6 \times 10^{41} {\rm GeV}^{-1}$ and $a_0 = 1$ (as given above) in (4.15), $t_*$ is calculated as
$$t_* = t_0 - 0.4 \times 10^{41}{\rm GeV}^{-1} = 6.2\times 10^{41}{\rm GeV}^{-1} = 12.86 {\rm Gyr}. \eqno(4.16)$$  

(4.14) shows that accelerated expansion (4.15) stops at $a = a_e$ satisfying
     the  condition
$$2.117 \times 10^{-47} a_e^{3/4} = 2 \lambda . \eqno(4.17)$$

$a(t)$, given by (4.15), acquires the value $a_e$ by the time
$$
t_e = t_* - 2 \times 10^{42} \Big[\sqrt{a_e^{-3/4} - \frac{2.117 \times
  10^{-47}}{2 \lambda}} - \sqrt{a_*^{-3/4} - \frac{2.117 \times
  10^{-47}}{2 \lambda}}  \Big] . \eqno(4.18)$$

The Friedmann equation (4.9) shows that, at $t \ge t_e$, expansion of the
universe is driven by matter again and it reduces to (4.12) yielding the
solution
$$a(t) = a_e [1 + 1.21\times 10^{-42} a_e^{-3/2}(t - t_e)]^{2/3} . \eqno(4.19)$$
which shows decelerated expansion. Thus another transition from {\em
  acceleration} to {\em deceleration} will take place at $t = t_e$.

It is interesting to note that the term 
$$ 2.117 \times 10^{-47} a^{3/4}
\Big[1 - \frac{2.117 \times 10^{-47} a^{3/4}}{2 \lambda} \Big]$$
will be negative as $ a > a_e$. So ,gradually universe will reach a state , where scale factor $a(t)$ acquires its maximum value $a_m$. At
$a = a_m, {\dot a} = 0$ in (4.9) and $a_m$ satisfies the condition
$$\frac{6.67 \times
  10^{-48}}{a_m^3} = 2.117 \times 10^{-47} a_m^{3/4} 
\Big\{\frac{2.117 \times 10^{-47} a_m^{3/4}}{2 \lambda} - 1\Big\}.
  \eqno(4.20)$$

$a_m$ gives the maximum expansion, so for $t > t_m$ universe will change its
direction and retrace back leading to contraction. As a consequence, for $t > t_m, a(t)$
will decrease and $a^{-3}$ term ,in (4.9), will dominate yielding the effective
equation 
$$ \frac{\dot a}{a} = - \frac{8.099 \times 10^{-43}}{a^{3/2}}. $$
Here, $\frac{\dot a}{a} < 0$ due to contraction. This equation is integrated
to
$$ a(t) = a_m [1 - 1.21 \times 10^{-42} a_m^{-3/2}(t - t_m) ]^{2/3}
. \eqno(4.21)$$  
(4.21) shows that at time
$$ t = t_m + 8.26 \times 10^{41} a_m^{3/2},  \eqno(4.22)$$

$a = 0.$ It means that universe will collapse at this time.

\bigskip

\centerline {\underline{\bf 5. Summary}}

\smallskip

 Results, obtained above, are summarized as follows. Here $f(R)-$
gravitational action is obtained by adding higher-order terms $R^2$ and $R^{(2
  + r)}$ of scalar curvature $R$ to the Einstein-Hilbert term. Gravitational
  field equations are derived from this action. Using $R \sim a^{-n}$ in
  trace of $f(R)-$ gravity field equation, Friedmann equation is obtained.  It
  is found above that if $r = 3$ and $n = 1/4$, FE (obtained here) contains
  quintessence like dark energy term as well as radiation like term in the
  early universe. Here,  radiation emerges spontaneously and is termed as dark
  radiation which is analogous to a similar term in brane-gravity-based
  FE. Dark energy term is induced by curvature and it vanishes when the scale
  factor $a(t)$ acquires a finite value $a_c$ given by (3.2b). It is
  interesting to see that, in the late universe, dark matter term emerges
  spontaneously and in the very late universe (when it is $12.86 {\rm Gyrs}$
  old) the universe is dominated by curvature-induced phantom dark energy with
  ${\rm w} = - 1.25.$  Contribution of $R$ to DE is a 
physical concept in addition to its usual role as a geometrical field. Thus, 
dual roles of $R$ (as a physical field as well as a geometrical
field)\cite{skp} are manifested here. The cosmological scenario, obtained
  here, from $f(R)-$ gravity with higher-order terms $R^2$ and $R^5$, is given
  as follows.

It is found that the early universe inflated for a very short period
with power-law speeded-up expansion, driven by curvature-induced quintessence
DE. When the scale factor increased upto $10^{28}$
times the scale factor at Planck scale, quintessence DE density vanished. As a
consequence, universe came out of the inflationary era. Later on, early
universe decelerated as $t^{1/2}$ driven by dark radiation. Subsequently, when
the universe became sufficiently old, it decelerated as $t^{2/3}$ driven by
dark matter. At red-shift $z_{**} = 0.3607$, a transition from deceleration to
acceleration took place and universe began to accelerate driven by
curvature-induced phantom DE explaining the present acceleration of the
universe. It is found that the late acceleration is {\em transient} and it
stops when phantom DE grows to a finite value equal to
$2\lambda$. Here $\lambda$ is the cosmic tension analogous to brane-tension,
which is explained above. Interestingly, the phantom model(obtained here) is free from the
menace of future-singularity. Here, it is shown that universe will reach its
maximum expansion at a time $t_m$. Later on, it will contract and collapse in
a finite future
time. Thus, it is found that contrary to $f(R)-$dark energy models,  we get
a viable cosmology from $f(R)-$ gravity. Also,  it is found that curvature induced phantom DE has a crucial role in the 
dynamics of present and future universe giving different phases mentioned
above. Results, obtained above, support observations made so far.


\bigskip


\begin{thebibliography}{14}

\smallskip

\bibitem{sp}
 S. J. Perlmutter $et$ $al.$, Astrophys. J. {\bf 517},(1999)565;
 astro-ph/9812133;  D. N. Spergel $et$ $al$,  Astrophys J. Suppl. {\bf 148}
 (2003)175[ astro-ph/0302209] and references therein.

\smallskip
\bibitem{ag}
 A. G. Riess $et$ $al$, Astrophys. J. {\bf 607}, (2004) 665 [
 astro-ph/0402512].

\smallskip

\bibitem{ejc}
 E.J.Copeland, M.Sami and S. Tsujikawa, Int. J. Mod. Phys. D, {\bf
 15},(2006)1753 [hep-th/0603057] and references therein. 

\smallskip

\bibitem{cap}
S. Capozziello, V.F.Cardone, S.Carloni and A.Troisi, Int. J. Mod. Phys. D, {\bf
 12},(2003)1969; S.M. Carroll, V.Duvvuri, M. Trodden and M.S.Turner,
Phys. Rev.D, {\bf  70},(2004) 043528.

\smallskip
\bibitem{snj}
S. Nojiri and S.D.Odintsov, Int.J. Geom. Meth. Mod. Phys. {\bf 4},(2007)115
[hep-th/0601213 ]and references therein.

\smallskip

\bibitem{lds}
L. Amendola, D. Polarski and S. Tsujikawa, Phys.Rev. Lett. {\bf 98} (2007)
131302 [astro-ph/0603703] ; L. Amendola, D. Polarski, R.Gannouji and S. Tsujikawa,  Phys.Rev.D,
{\bf 75} (2007) 083504 [gr-qc/0612180].

 




\smallskip
\bibitem{sks}
S.K.Srivastava, astro-ph/0511167;astro-ph/0602116; Int.J.Mod.Phys.A {\bf 22
  (6)} (2007), 1123-1134 [hep-th/0605019].
 

\smallskip
\bibitem{sks06}
S.K.Srivastava, Phys.Lett. {\bf B 643} (2006) 1-4 [astro-ph/0608241].

\smallskip
\bibitem{sks07}
S.K.Srivastava, Phys.Lett. {\bf B 648} (2007) 119-126 [astro-ph/0603601].




\smallskip
\bibitem{ms}
M. Sami, P. Singh and S. Tsujikawa, Phys. Rev.D, {\bf 74} (2006)043514[gr-qc/0605113].



\smallskip
\bibitem{rm}
 R. Maartens, gr-qc/0312059.




\smallskip
\bibitem{vs}
V. Sahni and Y. Shtanov, astro-ph/0202346.

\smallskip
\bibitem{bo}
B. Feng, X. Wang and X. Zhang, Phys. Lett. B, {\bf 607} (2005) 35 [astro-ph/0404224].

\smallskip
\bibitem{ad}
A.D. Miller $et$ $al$ , Astrophys. J. Lett. {\bf 524} (1999) L1; P. de
Bernadis $et$ $al$ , Nature (London){\bf 400} (2000) 955; A.E. Lange $et$ $al$
, Phys. Rev.D{\bf 63} (2001) 042001; A. Melchiorri $et$ $al$ ,
Astrophys. J. Lett. {\bf 536} (2000) L63; S. Hanay $et$ $al$ , 
Astrophys. J. Lett. {\bf 545} (2000) L5.


\smallskip
\bibitem{abl} 
A.B. Lahnas, N.E. Mavromatos and D.V. Nanopoulos,  Int. J. Mod. Phys. D, {\bf 
  12(9)}, 1529 (2003).  

\bibitem{skp}
 S.K.Srivastava and K.P.Sinha; Phys.Lett.B, {\bf 307} (1993) 40; 
  Pramana, {\bf 44} (1993) 333;  Jour. Ind. Math. Soc.{\bf 61},80 (1994); 
  Int.J.Theo.Phys., {\bf 35} (1996) 135;  Mod.Phys.Lett.A, {\bf 12} (1997) 
  2933; S.K.Srivastava; Il Nuovo  Cimento B, {\bf 113} (1998) 1239; 
  Int.J.Mod.Phys.A, {\bf 14} (1999) 875; Mod.Phys.Lett.A, {\bf 14} (1999)
  1021; Int.J.Mod.Phys.A, {\bf 15} (2000) 2917; Pramana, {\bf 60} (2003) 29; 
S.K.Srivastava, hep-th/0404170; gr-qc/0510086.  
\end{thebibliography}
\end{document}